\documentclass[jcp,twocolumn,superscriptaddress,floatfix,graphicx,showpacs]{revtex4-1}

\usepackage{ifpdf}
 \newif\ifpdf
\ifx\pdfoutput\undefined
   \pdffalse
\else
   \pdfoutput=1
   \pdftrue
\fi
\ifpdf
   \usepackage{graphicx}
   \usepackage{epstopdf}
\else
   \usepackage{graphicx}
\fi

\usepackage{srctex}
\usepackage{hyperref}

\usepackage{amsmath,amssymb}

\usepackage{epsfig}

\begin{document}


\title{Hydrodynamic anomalies in supercritical fluid}

\author{R.E. Ryltsev}
\affiliation{Institute of Metallurgy, Ural Division of Russian Academy of Sciences, 620016 Yekaterinburg, Russia}

\author{N.M. Chtchelkatchev}
\affiliation{Moscow Institute of Physics and Technology, 141700 Moscow, Russia}
\affiliation{Institute for High Pressure Physics, Russian Academy of Sciences, 142190 Troitsk, Russia}
\affiliation{L.D. Landau Institute for Theoretical Physics, Russian Academy of Sciences, 142432, Moscow Region, Chernogolovka, Russia}
\affiliation{Department of Physics and Astronomy, California State University Northridge, Northridge, CA 91330, USA}


\begin{abstract}
Using the molecular dynamics simulations we investigate properties of velocity autocorrelation function of Lennard-Jones fluid at long and intermediate time scales in wide ranges of temperature and density. We show that the amplitudes of the leading and subleading VAF time asymptotes, $a_1$ and $a_2$, show essentially non monotonous temperature and density dependence. There are two lines on temperature-density plain corresponding to maxima of $a_1$ ($a_2$) along isochors and isotherms situated in the supercritical fluid  (hydrodynamic anomalies). These lines give insight into the stages of the fluid evolution into gas.
\end{abstract}

\maketitle

\section{Introduction}

Dynamic correlation functions (DCF) are among the main resources of our information of condensed matter systems. Since the pioneering work of Alder and Wainwright~\cite{Alder1970PRA}, it was realized that long time  asymptotic behaviour of DCF in fluids may be essentially nonexponential due to collective particle motions. Now it is well established experimentally that in liquids, supercritical fluids, colloids, suspensions and many other soft condensed matter systems collective effects may give essential and even main contribution in DCFs at small frequencies (and momenta)~\cite{Kim1973PRL,Levesque1974PRL,Boon1975PhysLettA,Paul1981JPhysA,Morkel1987PhysRevLett,Zhu1992PRL,Kao1993PhysRevLett,Kulik2005PRL}. For velocity autocorrelation function (VAF) in fluid the leading long-time term $\sim a_1 t^{-\alpha_1}$, where $t$ is time and $\alpha_1=d/2$~\cite{comment1} with $d$ being dimensionality~\cite{Ernst1970PRL,Dorfman1970PRL,Grimmett1994,Cichocki1995PRE,Snook2001PRE,Ernst2005PRE,Dib2006PRE,Williams2006PRL}. In particular, for 2D fluid systems, $1/t$ in tail of VAF stands behind the divergence of the corresponding self-diffusion coefficient~\cite{Lowe1995PhysA,Schmidt2003JChemPhys,IsobePRE2008,Lin2014PRE}.

Naively, the tails of autocorrelation functions should monotonically decay as temperature increases (density decreases). But we find that tail amplitude $a_1$ demonstrates non monotonic behavior along isochors and isotherms. So there are two lines on temperature-density phase diagram which correspond to maximums on both $a_1(T)_{\rho=\mathrm{const}}$ and $a_1(\rho)_{T=\mathrm{const}}$ dependencies (see Fig.~\ref{Fig0} in Sec.~\ref{sec_discussion}).  On the set of such extrema we shall henceforth refer to as ``hydrodynamic anomalies''.

Important and nearly untouched problem is related to the behaviour of DCF at intermediate time (and spatial) scales where the hydrodynamic mode didn't come yet and also the one-partial description of VAF isn't already applicable. While $a_1$ and $\alpha_1$ can be obtained within linearised Navier-Stokes hydrodynamics, the subleading terms result from the generalised hydrodynamics beyond the Navier-Stokes limit. There is a conjecture that  the ``subleading'' term $a_2 t^{-\alpha_2}$ dominates in VAF at the intermediate scales, where in three dimensions, $\alpha_2=2-1/2^2=7/4$~\cite{Pomeau1973PRA,Ernst1975JPhys}. We confirm using molecular dynamics that indeed $\alpha_2\approx 7/4$ for Lennard-Jones (LJ) fluid and we find $a_2$. There is also hydrodynamic anomaly in $a_2(T,\rho)$, see Fig.~\ref{figtails}d.

Investigation even the amplitude $a_1$  requires much more accuracy compared with the power $\alpha_1$-law check up. Verification of the subleading power laws and understanding of $a_{n>1}$ behavior is even more challenging task. Physics of DCF tails is related to correlated move of large amount of particles. So its investigation requires, from one side, the huge amount of particles to avoid boundary effects and, on the other side, the unique calculating algorithms for storage and intense processing of the particle dynamic history at large time scales. As the result one can extract from DCF new information about the fluid system hardly available at small time scales.

Below we investigate VAF behavior at long and intermediate time scales in simple fluid in wide ranges of temperature and density. The basic fluid model used is the one component Lennard-Jones pair potential model. In order to study the influence of critical fluctuations and attractive part of the potential on VAF behavior, we checked the stability of our main results on the the soft-spheres model. Calculating VAF of the system with high accuracy we show that amplitude of VAF long-time tails demonstrates non trivial temperature and density dependence (hydrodynamic anomalies). We show that at intermediate time scales behavior VAF in fluid is out of the frameworks of linearised Navier-Stokes hydrodynamics and, moreover, changes qualitatively with temperature.

 \begin{figure}
  \centering
  \includegraphics[width=\columnwidth]{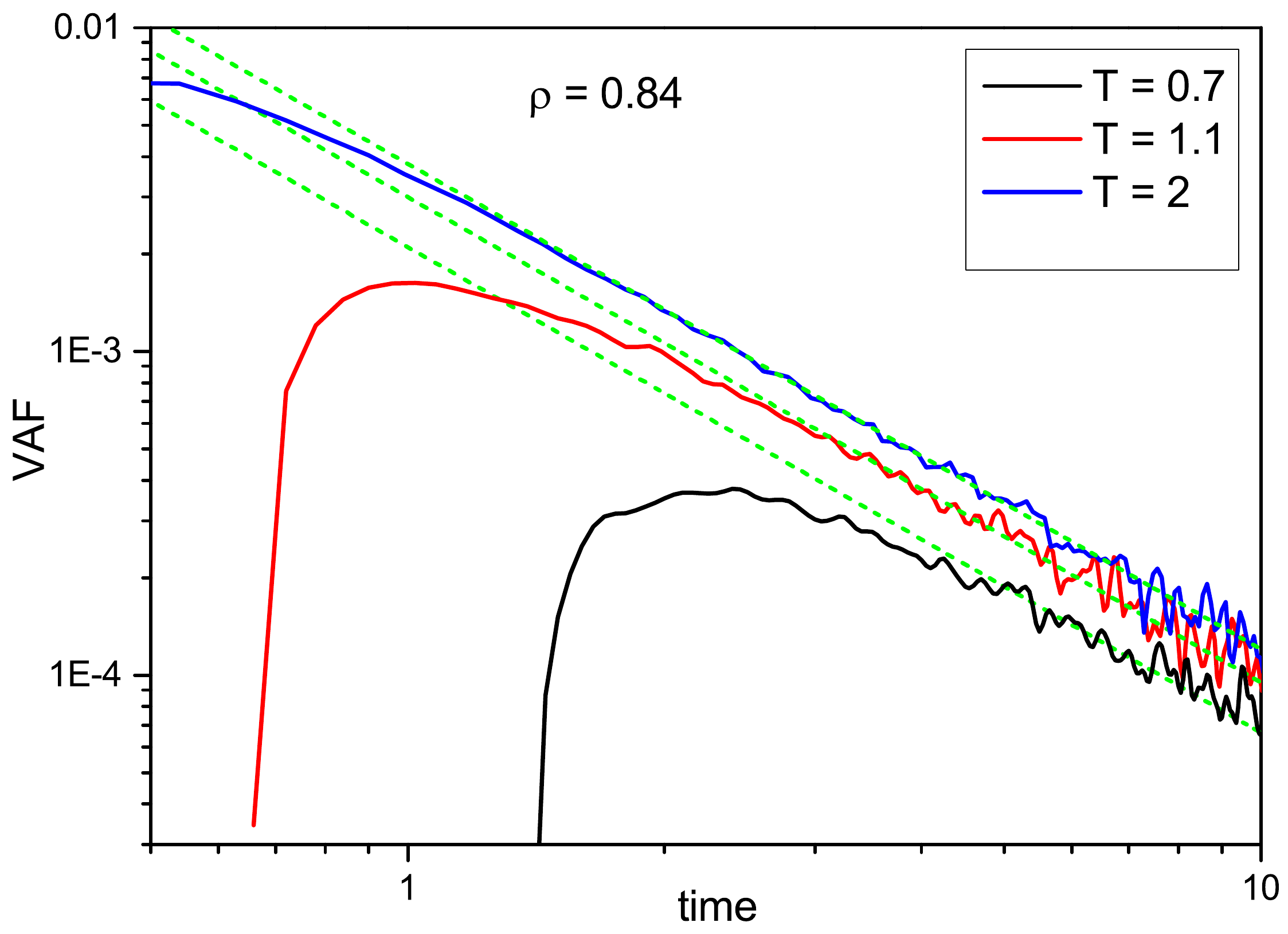}\\
    \caption{(Color online) $Z(t)$ at $\rho=0.84$ and different temperature. The straight dashed lines represent $a_1\,t^{-3/2}$ asymptotics with appropriate amplitude $a_1$. }\label{Fig1}
\end{figure}

 \begin{figure*}
  \centering
  \includegraphics[width=\textwidth]{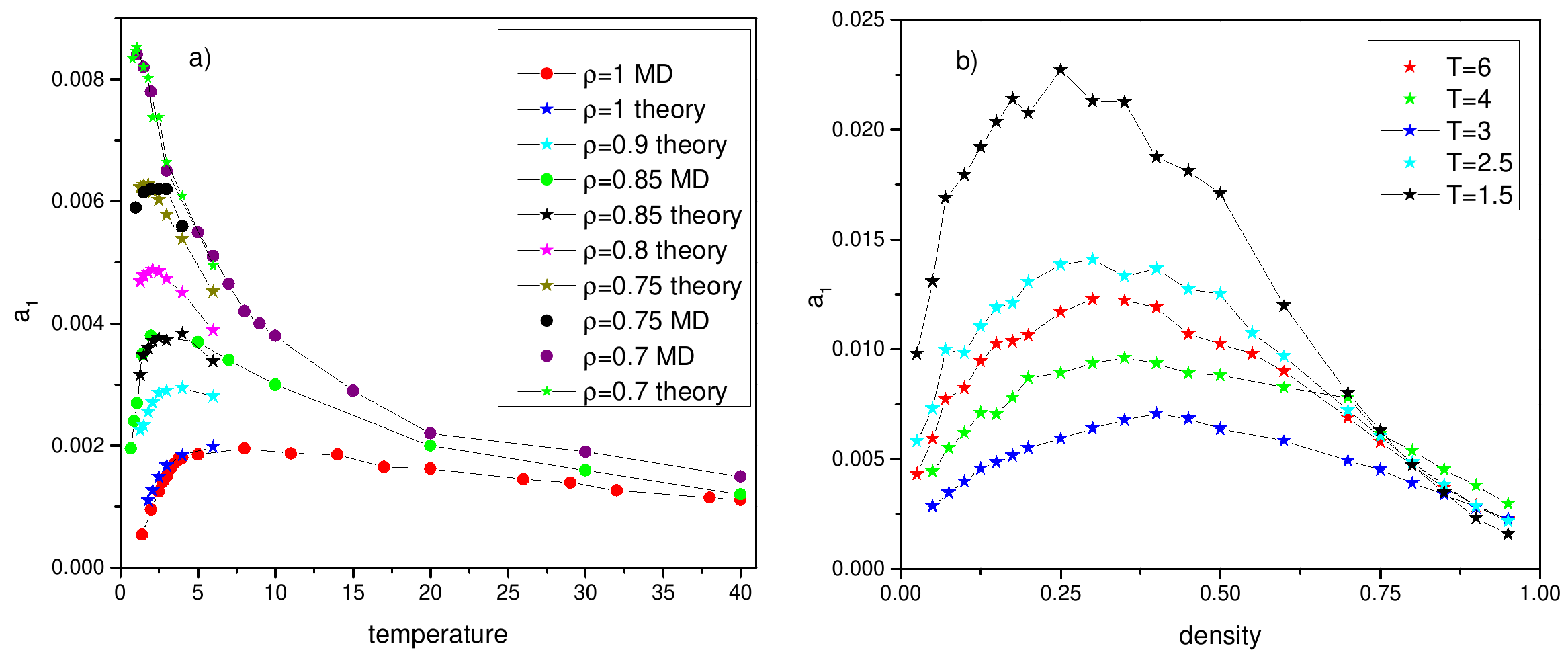}
  \caption{(Color online) The dependencies of VAF long-time tail amplitude $a_1$ for LJ fluid on temperature (a) and density (b). The bullets represent result of direct estimation from VAF obtained by molecular dynamic simulation and the stars are the results of calculations by using formula \ref{a1_theor}.}\label{a1(T)_a1(rho)}
\end{figure*}

\section{Simulation details}

To avoid the finite size effects at hydrodynamic scales one should consider large enough volume. The volume of fluid ``disturbed'' by the given moving particle grows as (see Sec.~8.7 in~\cite{hansenMcDonald}),
\begin{gather}\label{eq:Vh}
  V_h(t)\sim[t(\nu+D)]^{d/2}
\end{gather}
where $\nu$ is the kinematic viscosity. So the simulation volume $V$ should be much larger than $V_h(t_h)$, where $t_h$ is the characteristic time when the dynamic correlation functions start exhibiting the hydrodynamic asymptotic scaling. Generally, the minimum volume $V_h(t_h)$ is not universal: it depends on temperature and density.

It was recently shown in two-dimensional hard disk system at small and intermediate densities that reliable calculation of VAF hydrodynamic tails in fluid requires at least $64^2$ particles, see Fig.2 in Ref.~\cite{IsobePRE2008}. It was also shown that the insufficient volume results in not only strong increase of the calculation noise in the correlation function but also on the sudden suppression of $t^{-\alpha_1}$ hydrodynamic tail at time scales where $V_h(t)>V$.

We have considered the system of $N=128^3\approx2.1\times 10^6$ particles that were simulated under periodic boundary conditions in 3-dimensional cube in the Nose-Hover (NVT) ensemble. Even these amount of particles required terabytes of operational memory to process the particle trajectories and calculate VAF at satisfactory accuracy.  To optimize the calculation speed and the operational memory we developed the unique parallel algorithm that was built in the $\rm{DL\_POLY}$ Molecular Simulation Package~\cite{dl_poly,todorov2004dl_poly_3,todorov2006dl_poly_3}.

Simulating LJ particle system we apply pair potential in the standard form, $U(r)=4\varepsilon [(\sigma/r)^{12}-(\sigma/r)^{6}]$, where $\varepsilon$ – is the unit of energy, and $\sigma$ is the core diameter. In the remainder of this paper we use the dimensionless quantities for LJ fluid: $\tilde r= r/\sigma$, $\tilde U = U/\varepsilon$, temperature $\tilde T = T/\varepsilon$, density $\tilde{\rho}\equiv N \sigma^{3}/V$, and time $\tilde t=t/[\sigma\sqrt{m/\varepsilon}]$, where $m$ and $V$ are the molecular mass and system volume correspondingly. As we will only use these reduced variables, we omit the tildes. For the soft-sphere model we apply $U(r)=4\varepsilon (\sigma/r)^{12}$.

Note, that we will always use VAF normalized on its initial value
\begin{gather*}
Z(t) = \frac{{\langle {{\bf v}(0) \cdot {\bf v}(t)} \rangle }}{{\langle {\left| {{\bf v}(0)} \right|^2 } \rangle }},
\end{gather*}
where $\langle {\left| {{\bf v}(0)} \right|^2 }\rangle  = 3k_B T/m$.
\section{Results}

\begin{figure*}[t]
  \centering
  \includegraphics[width=\textwidth]{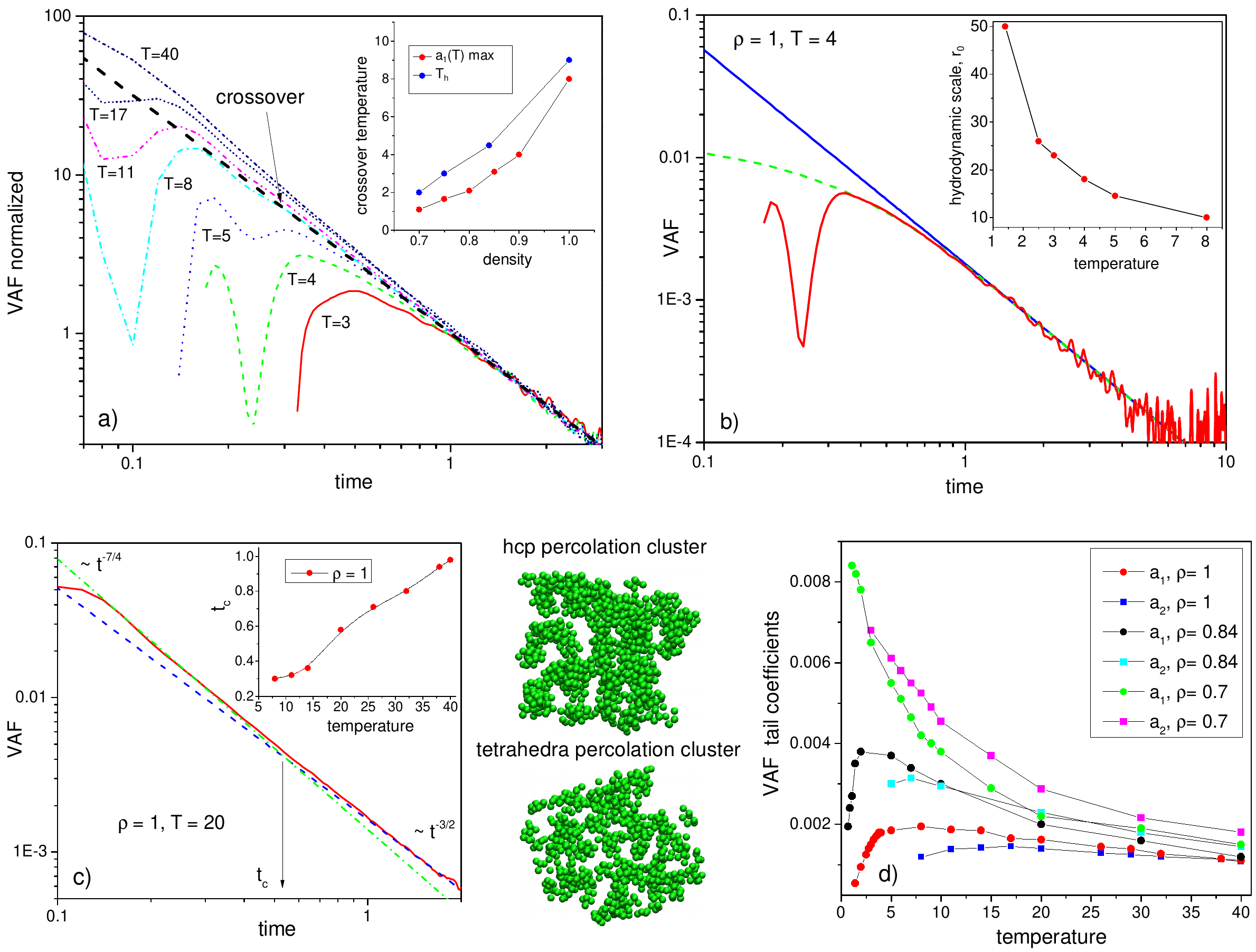}
    \caption{(Color online) (a) Master curve of VAFs normalised on corresponding $a_1(T)$ for $\rho=1$. The arrow shows VAF at crossover temperature $T_h$ at which the deviation of VAF from leading asymptotics changes its sign. The insert displays the dependence of crossover temperature $T_h$ on density. (b) Solid line represents VAF at $\rho=1$, $T=4$; straight line is the $t^{-3/2}$ asymptotics; dashed line is the integral (\ref{a1_theor}) with adjusted cutoff $r_0$. The insert shows the dependence of $r_0$ on temperature at $\rho=1$.(c) Solid line represents VAF at $\rho=1$, $T=20$;  dot-dashed and dashed straight lines are the subleading term $a_2t^{-7/4}$ and leading asymptotics $a_1t^{-3/2}$ respectively. The arrow shows the time $t_c$ corresponding to crossover to Navier-Stokes hydrodynamics regime. The dependence of this time on temperature at $\rho=1$ is shown in the insert. (d) Temperature dependencies of amplitude $a_2$ of subleading term corresponding to $n=2$ in the expansion (\ref{vaf_expansion}) in common with the same dependencies for leading term amplitude $a_1$. The snapshots demonstrate typical spatial distribution of atoms with hcp and tetrahedral local order at temperatures corresponding to their percolation thresholds.}\label{figtails}
\end{figure*}
We investigate LJ system in the density range of $\rho\in(0.7-2.0)$ at temperatures within the ranges from melting (or liquid-gas) line up to $T=300$. According to equilibrium temperature-density phase diagram \cite{smit1992JChemPhys, lin2003JChemPhys,khrapak2010PRB}, these ranges widely cover the region corresponding to both liquid and fluid states, see Fig.~\ref{Fig0}. At each thermodynamic point, we calculate VAF with high accuracy and analyze its behavior at intermediate and long time scales. First, we discuss long-time properties.

 \subsection{Long-time VAF behavior}

In Fig.\ref{Fig1}, we introduce typical $Z(t)$ curves at triple point density $\rho=0.84$ and different temperatures. It is clear from the picture that long-time VAF behavior is well defined by power function $a_1\,t^{-3/2}$ which is straight line in logarithmic scale used. We also see that the tail amplitude $a_1$ depends on the temperature and its change exceeds the error of VAF determination essentially. In Fig.\ref{a1(T)_a1(rho)}a the $a_1(T)$ dependencies at different densities are shown. We see that there are clear maxima on $a_1(T)$ curves in the density region $\rho\geq0.7$. The line joining these maxima on $T-\rho$ plain is drawn in Fig.~\ref{Fig0}. This line determines the states corresponding to anomaly of hydrodynamic VAF tails along isochors and so we will refer it to as isochoric hydrodynamic anomalies line.

The theoretical approach based on hydrodynamic approximation makes it possible to estimate VAF tail amplitude $a_1$. The final result is the following~\cite{hansenMcDonald,footnote0}:
\begin{equation}\label{a1_theor}
a_1 (T,\rho ) = \frac{{2}}{{3\rho}}\left[ {4\pi (D + \nu )} \right]^{ - 3/2}.
\end{equation}

We compared our numerical simulations of tail amplitude $a_1$ with formula (\ref{a1_theor}). We used the data of Meier \cite{meier2004JChemPhys, meier2004JChemPhys2} to get values of kinematic shear viscosity and self-diffusion coefficient; the values of $D$ were additionally calculated by Green-Kubo relation to control accuracy. The calculated tail amplitudes are in excellent agreement with those estimated from VAF directly (see Fig.~\ref{a1(T)_a1(rho)}) that provides the self-consistency of our results.

 Expression~(\ref{a1_theor}) appeared to be very accurate. That allowed us to extrapolate $a_1$ in the low-density region where VAF was not actually calculated within MD.
 Using the Meier data \cite{meier2004JChemPhys, meier2004JChemPhys2}, we calculated $a_1(T,\rho)$ at $\rho\in(0.02, 1)$, $T\in(1.5, 6)$. It was found that $a_1$ demonstrates areas with non-monotonic behavior along isotherms. In Fig.~\ref{a1(T)_a1(rho)}b we show $a_1(\rho)$ curves at different temperatures which reveal clear maxima. So there is one more hydrodynamic anomalies line, the isothermal one, which is the locus of $a_1$ extrema  obtained at isothermal conditions (see Fig.~\ref{Fig0}). Note that this anomaly can be also detected within the mode-coupling theory and it was seen in simulations of the lattice gas~\cite{Frenkel1991PhysicaD}.

 It should be noted that isochoric and isothermal hydrodynamic anomaly lines are located in essentially different areas of phase diagram.  The detailed discussion of physical meaning of these anomalies is presented in Section~\ref{sec_discussion}).

 \subsection{Intermediate-time VAF behavior}

It follows from our calculations that the long-time $t^{-3/2}$ behavior of VAF is ``universal''.  To demonstrate it, we introduce ``master curve'' which is a set of VAFs at different $T,\rho$ normalised on the corresponding $a_1(T,\rho)$ (see typical picture for $\rho=1$ in Fig.~\ref{figtails}a). We see universal asymptotics at $t\to\infty$. However the intermediate-time properties of VAF ``nonuniversally'' depend on both the temperature and the density at intermediate time scales: at $T=T_h$ ($T_h\approx 10$ for $\rho=1$)  the sign of VAF deviation from $t^{-3/2}$ changes. Here $T_h$ is certain characteristic temperature scale, investigated in detail below.

A deeper inside into nature of crossover temperature $T_h$ is afforded by considering the analytical expressions describing VAF at long times. It is established for simple liquids that the decay of the velocity autocorrelation function at large time scales is well described by the integral relation~\cite{hansenMcDonald}:
\begin{gather}\label{eqZ}
  Z_h(t,r_0)\sim \int_{k<2\pi/r_0}\exp{(-A k^2 t)}\frac{d^3k}{(2\pi)^3},
\end{gather}
originating from long-wave length hydrodynamics. Here  $A$ is estimated like $A\sim \nu+D$, where $D$ is the self-diffusion coefficient and $\nu$ is the kinematic shear viscosity. The cutoff $k_0=2\pi/r_0$ is related to the break up of hydrodynamic fluctuations at nanoscales $\sim r_0$. Typically $r_0$ is taken to be zero; then one can get only the leading asymptotic scaling, $\sim a_1 t^{-3/2}$ with $a_1$ defined by (\ref{a1_theor}). Here we keep finite $r_0$ that helps us to have progress in analytical description of VAF at intermediate time scales. In Fig.~\ref{figtails}b we show the fit of \eqref{eqZ} to VAF calculated using molecular dynamics. We see that adjusting the cutoff $r_0(\rho,T)$ we can describe VAF at long-time scales as well as at intermediate ones. It follows from (\ref{eqZ}) that $Z_h(t,r_0)\leq Z_h(t,0)\sim a_1 t^{-3/2}$. So keeping  finite  $r_0$ should always cause negative deviation of VAF from long-time asymptotics.

However this description breaks up at $T\approx T_h$. Insert in Fig.~\ref{figtails}b shows that $r_0$ (and so the negative deviation degree) decreases with temperature up to $T_h$ where $r_0$ is of the order of $l$, and $l$ is the average distance between molecules. At $T>T_h$ the fit \eqref{eqZ} strongly contradicts the simulation results at intermediate time scales: even the sign of the deviation (positive) cannot be reproduced. This contradiction is caused by the fact that at $T>T_h$ Eq.~\eqref{eqZ} based on purely Navier-Stokes hydrodynamics fails at intermediate time scales and so we have to use more rigorous expressions arising from generalized hydrodynamics.

It was shown within the frames of both the mode-coupling theory~\cite{Ernst1975JPhys} and the Enskog expansion~\cite{Pomeau1973PRA} that non-analytic dispersion relations for hydrodynamic frequencies may take place at intermediate time and spatial scales. That leads to infinite asymptotic expansion for time correlation functions. Particularly, for VAF that expansion has the form
\begin{gather}\label{vaf_expansion}
Z(t)\sim\sum_{n>0} a_n t^{1/2^n-2}.
\end{gather}

\begin{figure*}
  \centering
  \includegraphics[width=\textwidth]{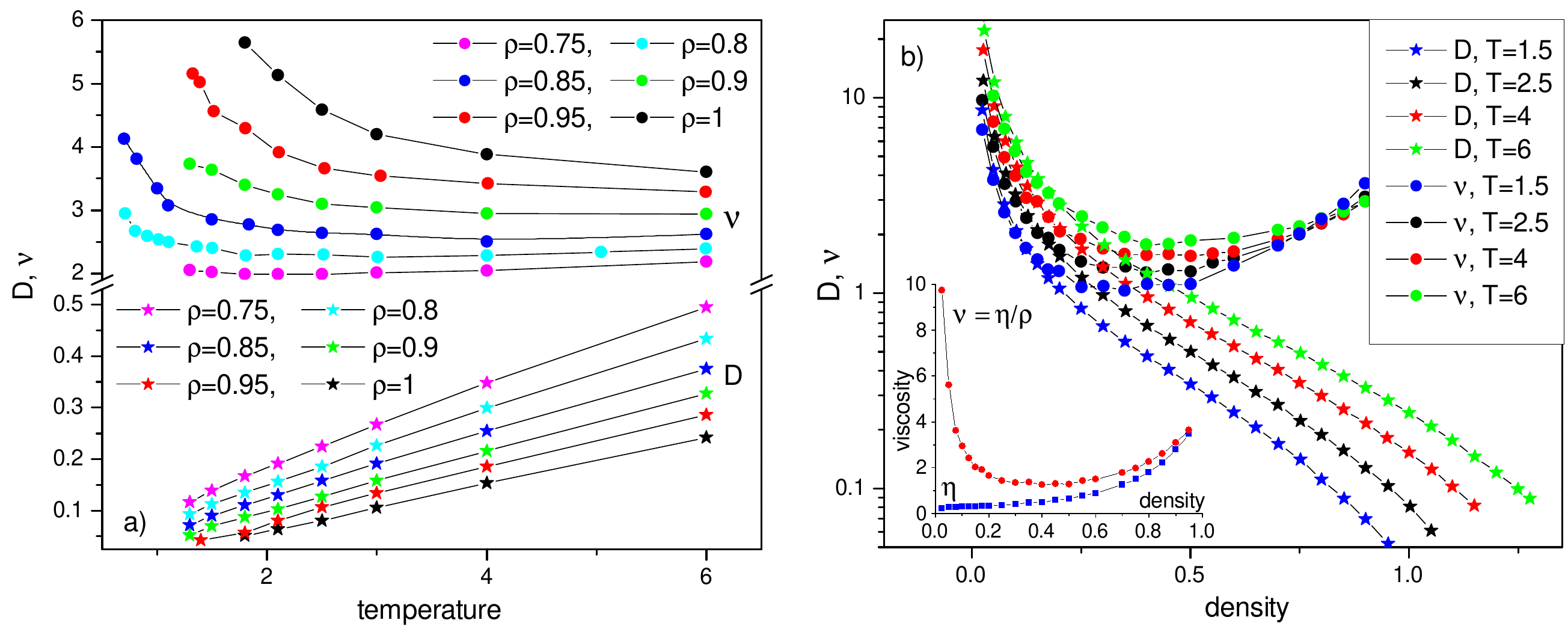}
  \caption{(Color online) The dependencies of self-diffusion coefficient $D$ and  kinematic shear viscosity $\nu$ on temperature (a) and density (b). The data were taken from \cite{meier2004JChemPhys, meier2004JChemPhys2}. The insert shows difference between density dependence of dynamic and kinematic viscosity. The discussion of this difference see in Sec.~\ref{SecViscos}}\label{Fig5}
\end{figure*}

The leading term of this expansion at $n=1$ is well known $a_1 t^{-3/2}$ and next subleading one is $a_2t^{-7/4}$. So it is natural to explain the positive deviation of VAF from $t^{-3/2}$ at intermediate time scales by the influence of these terms. In this connection, we note that the expansion (\ref{vaf_expansion}) is approximate one. Moreover this series diverges at $t\to\infty$. So we understand these series as just a hint that the term which behaves like $\propto t^{-7/4}$ may dominate at intermediate time scales. In Fig.~\ref{figtails}d, we see that the part of the VAF corresponding to intermediate time scales is well fitted by this term at $T>T_h$. The $t^{-3/2}$-behavior takes place at larger time scales and so we have a crossover to Navier-Stokes hydrodynamics characterized by a crossover time $t_c$ (see Fig.~\ref{figtails}c for explanation). This crossover time has predictable temperature dependence -- it increases with $T$ as we see in the insert of Fig.~\ref{figtails}c.

Thus our results reveal that intermediate-time VAF behavior at high enough ($T>T_h$) temperatures can be satisfactory described by $a_2t^{-7/4}$ term. So, adjusting the VAF by this term as pictured in Fig.~\ref{figtails}c, we can estimate its amplitude $a_2$. In Fig.~\ref{figtails}d, the typical temperature dependence $a_2(T)$ is shown in comparison with $a_1(T)$. We see that $a_2(T)$ is non-monotonic and so hydrodynamic anomaly also takes place.

It is important that crossover temperatures $T_h$ obtained at different densities are in close correlation with temperature of isochoric hydrodynamic anomaly. In the insert of Fig.~\ref{figtails}a, we see that these temperatures have almost the same density dependencies. Moreover, as can be seen from Fig.~\ref{figtails}d, the temperatures at which subleading terms are firstly detected correlate with those corresponding to $a_1(T)$ maxima.  This fact suggests the underlined physical reasons which cause the isochoric hydrodynamic anomaly and $T_h$ crossover are coupled. Of course, the additional study is needed to understand the physics of that effect completely but it is a matter for separate work.

\section{Discussion~\label{sec_discussion}}

In previous section we showed that amplitudes of VAF long-time tails demonstrate non monotonic behaviour which is expressed by the existence of isochoric and isothermal hydrodynamic anomalies lines. Moreover, the deviation of intermediate-time VAF from main $t^{-3/2}$ asymptotics changes its sign at crossover temperature $T_h$ which correlates to temperature of isochoric anomaly. The question arises that physics stands behind that behaviour? It is obvious that hydrodynamic anomalies revealed are related to some fundamental changes of fluid collective particle motion. Here we discuss these issues in more details.
\subsection{Hydrodynamic anomalies and viscosity~\label{SecViscos}}

We see in Fig.~\ref{a1(T)_a1(rho)} that $a_1$ are well described by (\ref{a1_theor}). According to this formula, tail amplitude behaviour is mostly defined by both the self-diffusion coefficient and kinematic shear viscosity. So one should look for physical explanation of the hydrodynamic anomalies in $D(T,\rho)$ and $\nu(T,\rho)$ dependencies. In Fig.~\ref{Fig5}, we show these dependencies for the areas of temperature-density phase diagram corresponding to the location of hydrodynamic anomalies.

First we focus on temperature dependencies of $\nu(T)$ and $D(T)$ (Fig.~\ref{Fig5}a) and investigate the nature of the isochoric hydrodynamic anomaly. We see that, at high density region where isochoric anomaly line takes place, the self diffusion coefficient is much smaller than the kinematic viscosity and so maxima on $a_1(T)$ dependencies are mostly induced by minima on $\nu(T)$ curves. The concurrent $D(T)$ contribution only shifts maxima of $a_1$ to smaller temperatures.

The very fact that temperature dependence of viscosity develops a minimum at high enough temperature is well known for a long time, see e.g. Refs.~\cite{meier2004JChemPhys,fomin2012JETPLett} and references therein. To understand this effect we should consider typical temperature dependencies of $\eta$ (or $\nu$) for dense fluids and gases. In liquids and dense fluids, shear viscosity decreases with temperature. The simplest model to show it is based on the assumption that the fluid flow obeys the Arrhenius equation for molecular kinetics. That gives exponential Arrhenius law $\eta\sim\exp(E/T)$ which works well for dense simple liquids and even for many liquid metals and alloys. In gases $\eta$ increases with temperature at fixed $\rho$~\cite{landau10}. In supercritical fluid, distinct liquid and gas phases do not exist -- this is some intermediate state and so there the liquid-like decrease of $\tau$ with temperature turns into the increase of $\eta$ with temperature like in gases, see review~\cite{Brazhkin2012UFNwanderings} and references therein.

\begin{figure}[t]
  \centering
  \includegraphics[width=\columnwidth]{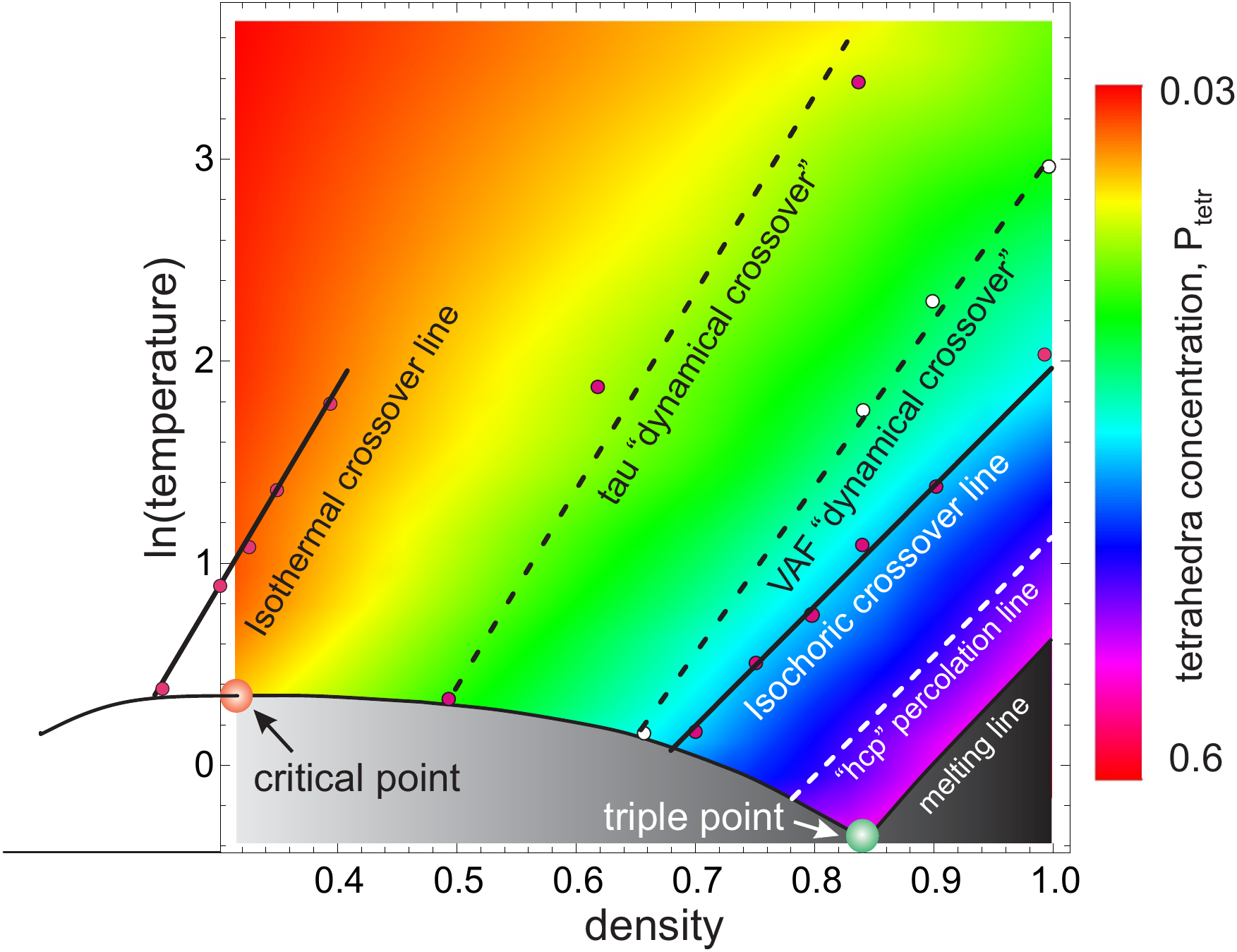}\\
    \caption{(Color online) The fragment of temperature-density phase diagram of LJ-system. The color gradients show concentration of tetrahedra clusters $P_{\rm tetr}$ extracted from \cite{Ryltsev2013PRE}. The solid lines show the positions of the isochoric and isothermal hydrodynamic anomalies corresponding to maxima of VAF long-time tails amplitudes along isochors and isotherms. Black dashed lines schematically show the dynamical crossover lines suggested in~\cite{brazhkin2012PRE, brazhkin2013PRL}. The white dashed line schematically locates the percolation line where the percolation cluster of the crystal-like hcp local order disappears. The percolation line of tetrahedral clusters approximately corresponds to the (hydrodynamic) isochoric crossover line.}\label{Fig0}
\end{figure}

Now, we turn to the density dependencies of $\nu(\rho)$ and $D(\rho)$ (Fig.~\ref{Fig5}b). We see that the self-diffusion coefficient decreases with density monotonously but viscosity has a minimum. This minimum stands behind the maximum on $a_1(\rho)$. It should be noted that the minimum on $\nu(\rho)$ curve is the spatial characteristic of kinematic viscosity; the dynamic viscosity increases with density monotonously (see the insert in Fig.~\ref{Fig5}b).

Fluid kinematic viscosity can be qualitatively identified with the resistance to flow and shear under the force of gravity. Here we assume the following gedanken experiment closely related to the method of capillary viscosimetry: taking the vertical open capillary and measuring the time $t$ of fluid downflow under the force of gravity one can find that $t\propto \nu=\eta/\rho$~\cite{landau2013fluid,wilke1994theory,miller1956JAP}.  At very low densities, $\nu$ has very large value -- dilute gas hardly flows through  capillary. That is why the kinematic viscosity diverges at $\rho\to 0$ whereas dynamic viscosity tends to zero. The decrease of $\nu$ in low density region is mostly due to the $\rho$ increase as far as $\eta$ is small. But, at intermediate densities, the grow of $\eta$ dominates and so the $\nu$ starts to increase developing the minimum. Thus we see that isothermal hydrodynamic anomaly is related to the crossover from dilute gas-like to fluid-like behavior in respect to flowing features. Note that this anomaly takes place at low density region.

\begin{figure}[t]
  \centering
  \includegraphics[width=0.8\columnwidth]{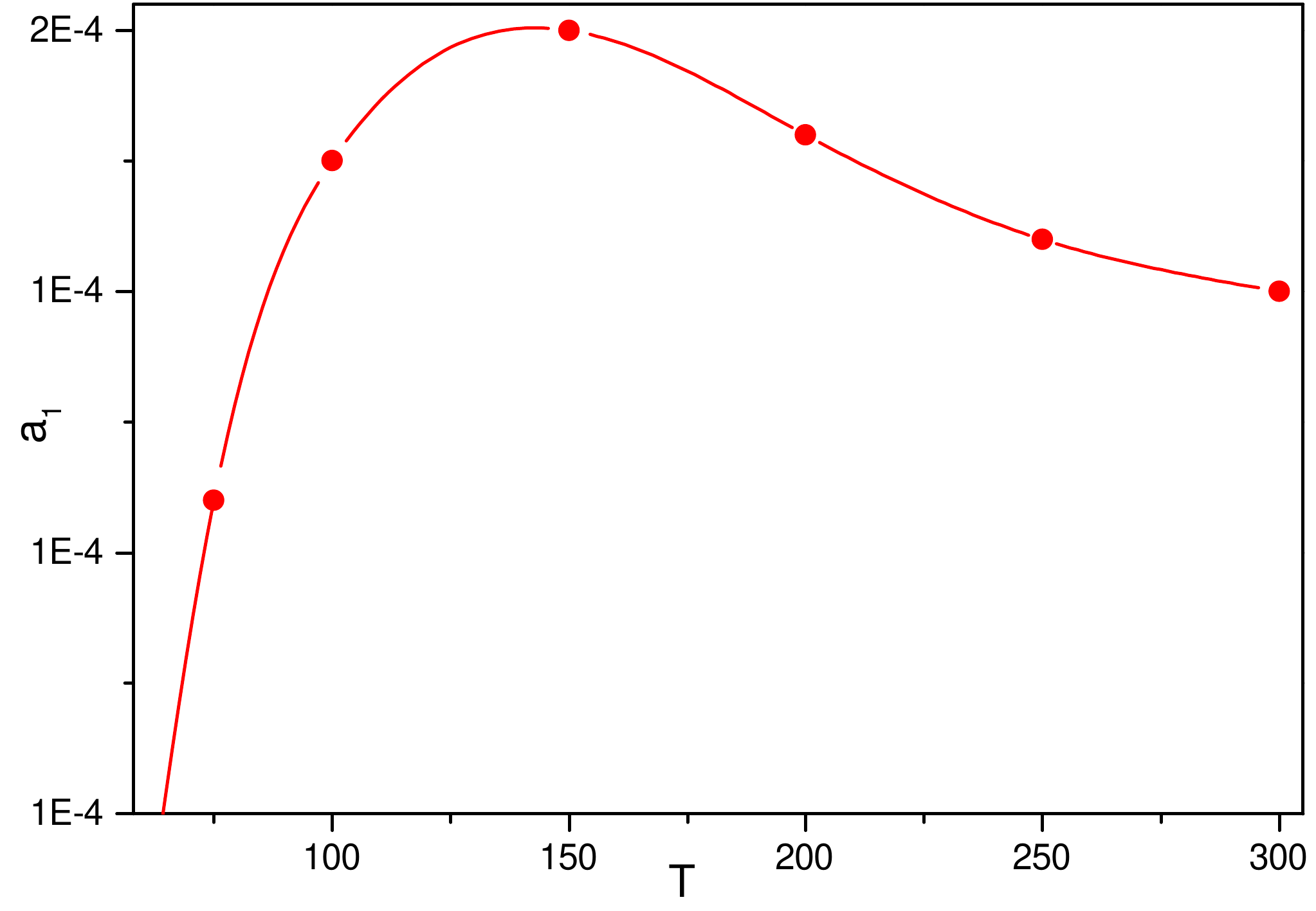}
    \\
  \caption{(Color online) Temperature dependence of long-time VAF tail amplitude $a_1$ at high density, $\rho=2$.} \label{Fig6}
\end{figure}

\subsection{Hydrodynamic anomalies and local structure evolution}

\begin{figure*}[t]
  \centering
  \includegraphics[width=\textwidth]{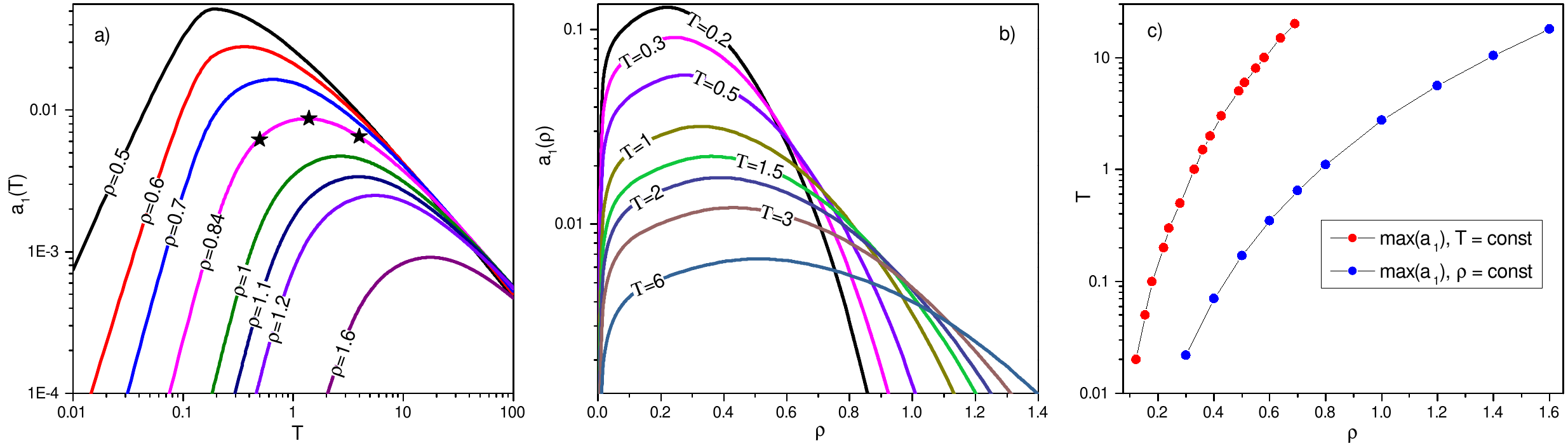}\\
    \caption{(Color online) The dependencies of tail amplitude $a_1$ for soft-spheres model on temperature (a) and density (b). The solid lines represent calculations by Eq.~\eqref{a1_theor} using polynomial fits for $D$ and $\nu$ from \cite{fomin2012JETPLett}. The stars are the results of direct evaluation of $a_1$ using MD. (c) The isochoric and isothermal hydrodynamic anomalies lines for soft-spheres model.}\label{fig:soft_sp}
\end{figure*}

It is well known that in liquids and fluids there is no long-range order but there is local order represented by molecule clusters with certain symmetries. The local structure demonstrates complex multistage evolution that changes with temperature and density. In particular, it was recently shown for simple fluids~\cite{Ryltsev2013PRE} that the solid-like close packed (hcp,fcc) clusters disappear at temperatures several times higher the melting temperature. But, surprisingly, the fragments of the close packed clusters -- tetrahedra -- survive at temperatures up to the several orders of magnitude higher than the melting temperature. This is shown in Fig.~\ref{Fig0} by the color gradient that corresponds to tetrahedra fraction.

The elements of local order may connect to each other and form the large-scale clusters that may in turn percolate. So, effectively, the local order may be the collective phenomena at certain conditions. Thus, there is a reason to search for the possible interference between the large-scale clusters and the long-time behaviour of dynamic correlation function that also corresponds to some collective phenomena. Our working hypothesis is that the local order and the (hydro)dynamical lines are somehow coupled with each other. The background of this hypothesis is the evident correspondence of the tetrahedra local order percolation line with the isochoric hydrodynamic anomaly. Of course, there is a big question why we see no percolation near by the isothermal line and this is still an open interesting issue.

To be more specific, we approximately locate areas of the phase diagram where the percolation clusters formed by the mentioned local order elements disappear. To locate the percolation threshold we used quite raw method: we visually examined the snapshots (see inserts in Fig.~4). The percolation line where the cluster of the tetrahedral local order disappears approximately corresponds to the (hydrodynamic) isochoric crossover line in Fig.~\ref{Fig0}.
 
The percolation line of crystal-like (hcp) clusters is predictably located near the melting line (see Fig.~\ref{Fig0}). This line approximately shows the threshold where the fluid loses symmetry properties inherited from the crystal phase. It is interesting question if any dynamical collective effect is related to this line. We remind that the stable low-temperature crystal phase of LJ system is fcc but the supercooled liquid has mostly hcp local symmetry.
 Metastable polycrystal forming during crystallization of the liquid is ususally the ``mixture'' of hcp and fcc clusters~\cite{Barron1955ProcRoySocLon}.
  
Of course we should emphasize that the detailed investigation of the percolation problem requires much more rigorous methods like the ``Swiss cheese'' percolation model~\cite{Lorenz2001JCP}. This is the matter of separate investigation.

\subsection{Hydrodynamic anomalies and ``dynamical crossover''}
We must note that minimum on $\nu(T)$ (or $\eta(T)$) has already been discussed in the context of so called ``dynamical crossover line'' [``crossover'' from liquid-like to gas-like behaviour in fluid]. It was written that minimum viscosity locus should coincide with dynamical crossover line defined by the criterion, $\tau\approx\tau_0$ where $\tau$ is the liquid relaxation time and $\tau_0$ is the minimal period of transverse quasiharmonic waves~\cite{brazhkin2012PRE,brazhkin2012JETPLett2}. Despite of qualitative soundness of the arguments presented, quantitatively, these lines are located far away from each other (see tau dynamical crossover line in Fig.~\ref{Fig0}). Later, the same authors proposed another criterium for dynamical crossover line based on features of VAF at atomic time scales~\cite{brazhkin2013PRL}. This ``VAF dynamical crossover line'' joins state points where oscillations of VAF disappear. Similar arguments have been used in the textbook~\cite{hansenMcDonald}, see, e.g., Chapter~7, where the disappearance of VAF oscillations in fluid have been discussed. In particular, in Ref.~\cite{hansenMcDonald}, the criterium, $\Omega\tau\approx1/2$, for the dynamical VAF crossover has been derived, where $\Omega$ is the Einstein frequency and $\tau$ is the relaxation time standing in the self-diffusion coefficient, see, e.g., Eq.~7.3.30 in Ref.~\cite{hansenMcDonald} and also Ref.~\cite{Rice1966JChemPhys}.

It should be stressed, however, that these criteria deal with VAF features at small times and they are quite different from hydrodynamic anomaly line reflecting the long-time behaviour. The locations of these lines are also different (see VAF dynamical crossover line in Fig.~\ref{Fig0}). Frankly speaking, the location of some transition line from short-time behaviour of VAF is quite questionable. The criterium of VAF oscillation degradation is an ambiguous one: there is no much sense in oscillations when their frequency is of the same order as damping. Thereupon more reasonable to deal with criterium that is related to disappearance of negative part of VAF at small time scales. This negative part is related to backscattering of particles in ``cage'' of the surrounding particles~\cite{hansenMcDonald}. [This is different from the criterium that we discuss below in Sec.~\ref{SecManyfold} related to backscattering extinction in long time asymptote of VAF that exactly locates the melting line.] But this criterium will produce another line in the phase diagram different from the known ones. By the way, this line is very close to the isochoric hydrodynamic anomaly line shown in Fig.~\ref{Fig0}. All that suggests that there is no universal criterium describing the evolution of supercritical fluid system from liquid to gas~\cite{Ryltsev2013PRE}. Each property evolves with temperature and density in various way and so there is a continuous multistage process of transformation from liquid-like to gas-like behaviour.

\subsection{Stability of hydrodynamic anomalies}
\subsubsection{Stability of hydrodynamic anomalies to high density conditions}
We investigated hydrodynamic anomalies at wide enough density and temperature ranges within LJ-model. However there is a natural question how important the attraction part of the LJ-potential is. In order to check this point, we investigated VAF behaviour at high enough density, $\rho=2$. The result is shown in Fig.~\ref{Fig6}. We see that there is well defined maximum of $a_1(T)$ and so we can conclude that the hydrodynamic anomaly survives at high densities.

\subsubsection{Hydrodynamic anomalies in soft-sphere model}
In order to test our main results, we performed simulation of VAF for the soft-spheres (SS) model. There is no critical point in this model that allows to exclude influence of critical fluctuations on hydrodynamic anomalies. Moreover, due to rigorous scaling relations, it is possible to develop analytical approximations for transport coefficients~\cite{Hoover1970JChemPhys,Heyes2009MolPhys,fomin2012JETPLett} (and so for $a_1$). We calculate tail amplitude $a_1$ for SS by Eq.~\eqref{a1_theor} using polynomial fits for $D(T,\rho)$ and $\nu(T,\rho)$ from \cite{fomin2012JETPLett} and compared analytical calculations with MD results for one density and three values of temperature. The results show good agreement (see Fig.~\ref{fig:soft_sp}). We have seen above that Eq.~\eqref{a1_theor} works well for LJ in wide range of temperature and density. We believe now that Eq.~\eqref{a1_theor} works as well for SS. So we rely on the analytical expressions for $a_1(T,\rho)$ investigating hydrodynamic anomalies in SS.

In Fig.~\ref{fig:soft_sp}a,b, we show the dependencies of tail amplitude $a_1$ for soft spheres model on temperature (a) and density (b) correspondingly. We see that $a_1(T)$ and $a_1(\rho)$ dependencies are qualitatively similar to those for LJ model. Both the isochoric and isothermal hydrodynamic anomalies also take places. Corresponding anomaly lines are shown in Fig.~\ref{fig:soft_sp}c and demonstrate LJ-like behavior at intermediate and high densities.  At low-density and low-temperature region, the anomaly lines for LJ system are buried under liquid-gas cupola and so hardly available in MD simulation.

\paragraph*{To summarise,} the attraction part of the interaction potential is not very important for formation of hydrodynamic anomalies and  obviously, the existence of the hydrodynamic anomalies is not related to the influence of the critical point.

\subsection{The manifold of hydrodynamic anomalies~\label{SecManyfold}}
We should note here that the locations of any anomaly lines determined from the local extrema of thermodynamic functions are different depending on the trajectory in the space of the thermodynamic parameters. If we consider, for example, the isobar or isoentropy paths in the thermodynamic parameters space, we will get two additional lines, located in two different positions.

In a similar way one can consider the long time tails of other Green-Kubo correlators, e.g., the stress tensor correlator or correlator of the the energy flux, which also reveal nonexponential time decay~\cite{Pomeau1975PhysRep}. There are analytical expressions for the amplitudes standing by the leading long-time terms and these expressions are quite different from Eq.~\eqref{a1_theor}. So on can expect that the locations of the hydrodynamic anomaly lines (if any) differ from VAF hydrodynamic anomaly lines.

The natural question arises if there is an anomaly of a time correlator that does not depend on the path in the thermodynamic parameters space. This is so for the sign of VAF long time amplitude $a_1$ and the value of the time tail power $\alpha_1$. As far as we know, $a_1$ changes its sign when path in the thermodynamic parameters space crosses the melting line (positive in the fluid, negative in the supercooled region) and at the same time $\alpha_1$ changes stepwise from $3/2$ to $5/2$~\cite{Williams2006PRL}. Our MD calculations confirm this well known result.


Finally, we should emphasize again that the hydrodynamical lines should not be associated to some ``phase transition'' or ``narrow crossover''. There is a conjecture that the evolution of the supercritical fluid towards the ideal gas is the continuous multistage process where one liquid properties disappear earlier (or later) than others. This is so for the local structure evolution as we have demonstrated in Ref.~\cite{Ryltsev2013PRE}.

The wide spread of the hydrodynamic anomaly lines over the phase diagram, see Fig.~\ref{Fig0}, and the relatively large width of the hydrodynamic anomaly maxima in $a_1$ and $a_2$, see Figs.~\ref{a1(T)_a1(rho)}, \ref{figtails} and~\ref{fig:soft_sp}, are another arguments about absence of narrow crossover between the liquid-like and gas like fluid and this is one more evidence of the continuous multistage nature of fluid evolution to gas.

\section{Conclusions}

In conclusion, we investigate fitches of  dynamical velocity correlation function at long-time scales and verify analytical expression for the tail amplitude $a_1$ standing by the leading time asymptotics $t^{-3/2}$. Taking the Lennard-Jones fluid as an exemplary system we show that long-time and intermediate-time behavior of VAF changes essentially depending on the temperature and density. In particular, the lead amplitude $a_1$ has non-monotonous behaviour (see Fig.~\ref{a1(T)_a1(rho)}).  Two lines on temperature-density plain correspond to maxima of $a_1$: isochoric and isothermal extremuma (see Fig.~\ref{Fig0}). These hydrodynamic anomaly lines separate regions of phase diagram where fluid shows different behaviour related to collective particle motion. The comparison of our results with data obtained earlier suggests that there is no universal criterium describing the evolution of supercritical fluid system from liquid to gas. Each property evolves with temperature and density in various way and so there is a continuous multistage process of transformation from liquid-like to gas-like behaviour.

\section{acknowledgments}

This work was supported by Russian Scientific Foundation (grant RNF №14-12-01185). We are grateful to Supercomputer Center of Ural Branch of Russian Academy of Sciences for access to ``Uran'' cluster.

\bibliography{our_bib}

\begin{thebibliography}{53}
\bibitem [{foo()}]{footnote0}%
  \BibitemOpen
  \href@noop {} {\bibinfo  {journal} {Mean square displacement of particles in
  fluid and VAF are closely related. So the estimate~\eqref{eq:Vh} for $V_h$
  formally follows from Eq.~\eqref{a1_theor}, see, e.g., for review Sec.~8.7 in
  Ref.~\cite{hansenMcDonald}}\ }\BibitemShut {NoStop}%
\end{thebibliography}

\end{document}